\documentclass[12pt]{article}
\usepackage{amssymb,amsmath}
\usepackage{natbib}
\usepackage{graphicx}
\usepackage{epsfig}

\begin{document}

\title{On the recurrence and Lyapunov time scales of the motion
near the chaos border}
\author{Ivan~I.~Shevchenko \\
Institute of Theoretical Astronomy, \\
Russian Academy of Sciences \\
Nab.~Kutuzova~10, St.Petersburg 191187, Russia}

\date{}

\maketitle

\abstract{Conditions for the emergence of a statistical
relationship between $T_r$, the chaotic transport (recurrence)
time, and $T_L$, the local Lyapunov time (the inverse of the
numerically measured largest Lyapunov characteristic exponent),
are considered for the motion inside the chaotic layer around the
separatrix of a nonlinear resonance. When numerical values of the
Lyapunov exponents are measured on a time interval not greater
than $T_r$, the relationship is shown to resemble the quadratic
one. This tentatively explains numerical results presented in the
literature.}
%Refs.~\cite{SFL90,LFM92,MLF94,LD93}.}

%PACS numbers: 03.20, 05.45, 05.47, 46.10, 95.10C, 96.30H.

Key words: resonance, chaos, Hamiltonian system, Lyapunov exponents,
asteroid.

\section{Introduction}

On the basis of a lot of numeric experiments in problems dealing
with the dynamics of objects of the Solar system, Soper et
al.~\cite{SFL90}, Lecar et al.~\cite{LFM92} and Murison et
al.~\cite{MLF94} argued that the times of ``sudden changes'' in
the chaotic orbital behaviour could be statistically predicted by
means of computation of the largest Lyapunov characteristic
exponents (LLCE). They established that a simple ``universal''
statistical dependence existed between the time of a sudden
orbital change (designated henceforth $T_r$, the ``recurrence
time'') and the Lyapunov time (the inverse of the numerically
measured LLCE): $T_r \propto T_L^\beta$, with a typical value of
$\beta = 1.7~\div 1.8$; though a considerable dispersion of the
statistical data was usually present. The same kind of dependence,
with $\beta \approx 1.9$, was found by Levison and
Duncan~\cite{LD93} in simulations of the dynamics of the outer
Solar system, namely, the Kuiper asteroidal belt.

Morbidelli and Froeschl\'e~\cite{MF96} recently considered reasons
for the appearance of a statistical relationship between Lyapunov
times and ``macroscopic diffusion'' times in general nearly
integrable Hamiltonian systems. In particular, they showed
qualitatively that in the regime of multiple resonance overlapping
the relationship should be polynomial. However, they did not find
any theoretical indication for a universal power law, i.e. a power
law with a particular preferable value of the exponent. In their
analysis, they considered the strength of perturbation as a main
governing parameter, through which both times of interest were
expressed. They did not consider any numeric effects; in
particular, Lyapunov exponents were treated to be equal to their
theoretical values defined on the infinite time scale.

In what follows, a different approach is used. I study a role of
the choice of the starting values of trajectories, but not the
role of the values of parameters. This implies consideration of
selection effects arising in numeric computations due to natural
time limitations: of course, the LLCE of a chaotic trajectory
computed on the infinite time scale should not depend on starting
values, if these starting data belong to one and the same
connected chaotic region of phase space. So, a ``local'' numeric
definition is used in what follows for the LLCE. This means that,
since the LLCE is computed numerically on a finite time scale, a
chaotic trajectory may explore only a finite local domain of all
available chaotic region, but not necessarily all this region.

I show that, due to the basic phenomenon of the divided phase space,
or, in other words, due to the phenomenon of the presence of
chaos border, and the immanent sticking behaviour of chaotic
trajectories, there emerges a generic statistical dependence
between the times of chaotic transport (recurrence times)
and Lyapunov times, when a natural time
limitation is used in numeric simulations. This limitation is that
the LLCE are computed on time scales equal to or less than
the recurrence times. The derived generic relationship
explains tentatively the cited results~\cite{SFL90,LFM92,MLF94,LD93}.

Besides, two new numeric examples are given demonstrating the emergence
of the generic relationship straightforwardly. One of them concerns
the separatrix map describing the motion near the separatrix of a
general nonlinear resonance; the second one deals with the asteroidal
motion in the 3/1 mean motion commensurability with Jupiter.

\section{Computing Lyapunov exponents}
\label{comp}

Remember~\cite{LL92} that the LCE of an orbit measures the rate
of exponential divergence of trajectories close to this orbit
(about the general definition of the LCE of a function
see e.g. Ref.~\cite{A92}).
Let $d(t_0) \ll 1$ be the initial displacement of a shadow trajectory
from the main one, and $d(t)$ be the displacement at time $t$.
Then the LCE is determined by the formula

\begin{equation}
\sigma= \limsup_{{t \to \infty} \atop
         {d(t_0) \to 0}} {1 \over {t-t_0}} \ln{d(t) \over d(t_0)}.
\label{LLCE}
\end{equation}

\noindent Depending on the direction of the initial displacement
in phase space, the quantity $\sigma$ of a trajectory of a
Hamiltonian system can have $2N$ generally different values, where
$N$ is the number of degrees of freedom. However, on an everywhere
dense set of starting values of shadow trajectories, it attains
the single (maximum) value, the ``largest LCE'', shortly, LLCE.

Numerically, the LLCE is found, of course, on a finite time interval
(say, of $M$ time units) by means of the formula

\begin{equation}
\sigma(M)={1 \over {M \Delta t}}\sum_{i=1}^M \ln{r_i},
\label{NLLCE}
\end{equation}

\noindent
where $r_i$ is the ratio of the current displacement (at $t = i$)
to the preceding one (at $t = i-1$),
$\Delta t$ is the size of the time unit. Current displacements
should be periodically renormalized to some small value
(preserving direction of the displacement)~\cite{LL92},
so that the shadow trajectory is kept in a vicinity
of the main one.

The traditional numeric procedure of computation of the LLCE is:
build the dependence $\log \sigma(M)$, given by Eq.~(\ref{NLLCE}),
versus $\log M$, and find the value of $\log \sigma$ at which the
dependence is ``saturated'', i.e. attains a form of a horisontal
plateau (cf. e.g. Ref.~\cite{W83}). Before saturation, $\log
\sigma$ goes down on average linearly with $\log M$.

So, on one hand, the value of the LLCE is computed on time intervals
not less than the time of saturation; on the other hand,
the time of computation cannot be infinite. These limitations
from below and above impose certain selection effects
which are necessary to take into account in any numeric statistical
study incorporating LLCE.

\section{The separatrix map and the generic relationship}
\label{smgr}

The nonlinear pendulum provides a model of a nonlinear resonance
under very general conditions~\cite{C79,LL92}. The motion in a
vicinity of the separatrix of the nonlinear pendulum (the
nonlinear resonance) is described by the separatrix
map~\cite{C79,LL92}, hereafter SM. In what follows the SM in
Chirikov's form~\cite{C79}, or, in identical terms, the ``whisker
map'' is used. It was deduced by Chirikov~\cite{C79} for the
Hamiltonian

\begin{equation}
H = H_0 + \Lambda \cdot (\xi_1 + \xi_2),
\label{hp}
\end{equation}

\noindent
where $H_0 = {{{\cal G} p^2} \over 2} - {\cal F} \cos \kappa$
is the Hamiltonian of the pendulum, and the perturbation is
given by the terms
$\xi_1 = \cos (\kappa - \sigma)$,
$\xi_2 = \cos (\kappa + \sigma)$
(where $\sigma = \Omega t + \sigma_0$)
and is thus periodic and symmetric.
In what follows, the angle $\kappa$ is referred to as the
pendulum's angle, and $\sigma$ as the phase angle of perturbation.
The quantity $\Omega$ is the perturbation frequency, and
$\sigma_0$ is the initial phase; $p$ is the momentum;
${\cal F}$, ${\cal G}$, $\Lambda$ are constants.

The SM is a two-dimensional area-preserving map,
with an action-like and phase-like variables, the former one
measuring the relative deviation of the energy of the pendulum
(with respect to the unperturbed separatrix value),
and the latter one measuring the phase of perturbation.
The SM in Chirikov's~\cite{C79} form is

\begin{eqnarray}
w_{i+1} & = & w_i - W \sin \sigma_i,  \nonumber \\
\sigma_{i+1} & = & \sigma_i +
                   \lambda \ln {32 \over \vert w_{i+1} \vert}
                   \ \ \ (\mbox{mod } 2 \pi),
\label{sm}
\end{eqnarray}

\noindent
where $w$ denotes the relative pendulum's energy:
$w = {H_0 \over {\cal F}} - 1$. Constants $\lambda$ and $W$ are
parameters: $\lambda$ is the ratio of $\Omega$, the perturbation
frequency, to $\omega_0 = ({\cal F G})^{1/2}$,
the frequency of small-amplitude pendulum oscillations; and

\begin{equation}
W = {\Lambda \over {\cal F}} \lambda
\left( A_2(\lambda) + A_2(-\lambda) \right) =
{\Lambda \over {\cal F}}
{4 \pi \lambda^2 \over \sinh {\pi \lambda \over 2}},
\label{W}
\end{equation}

\noindent
where the function $A_2(x)$ denotes values of the Melnikov--Arnold
integral as defined in~\cite{C79}.
One iteration of the SM corresponds to one period of
pendulum's rotation or a half-period of its libration.

The SM is given by the same Eq.~(\ref{sm}) for many other kinds
of perturbation terms as well (of course, formulae for the parameter
$W$ are then different; see a particular example in Ref.~\cite{ZEA91}).
So, the SM provides a very general description of the motion
in a vicinity of the separatrix of a nonlinear resonance.

The SM can be locally linearized in $w$ to give the standard
map~\cite{C79}. In other words, the chaotic layer is locally described
by the standard map. Consider the chaotic motion in a vicinity of the
critical curve separating regular and chaotic domains.
Following designations of Refs.~\cite{CS86,C90}, let
$r_n = p_n/q_n$ be the continued fraction convergents to the winding
number of the critical curve. They represent the winding numbers of
principal resonances close to the critical curve. The stability of a
periodic trajectory $r_n$ can be characterized by the value of the
Greene residue~\cite{G68,G79}. Its value for a principal resonance
close to the critical curve is given~\cite{CS86} by the formula

\begin{equation}
R_n = R^{(1)} \exp \left( 1.20 q_n^{1+\varepsilon} (K - K_G) \right),
\label{Rn}
\end{equation}

\noindent
where $K$ is the stochasticity parameter of the approximating standard
map, $K_G \approx 1$ is its critical value, $R^{(1)} \approx 1/4$ is
the critical value of the Greene residue; $\varepsilon = 0.013$.
Eq.~(\ref{Rn}) follows from Greene's relation
$R \propto K^q$~\cite{G79} applied at chaos border, though it is
somewhat modified: the coefficient $1.20$, put instead of $1$, is
empirical.

As Chirikov and Shepelyansky~\cite{CS86} noted, Eq.~(\ref{Rn}) has
simple physical meaning: the locally defined Lyapunov exponents

\begin{equation}
l_n \approx {\ln (4 R_n) \over q_n}
\label{lan1}
\end{equation}

\noindent
practically do not depend on $q_n$, i.e. on a particular trajectory,
and are equal to the locally defined Kolmogorov--Sinai
entropy $h \propto \Delta K \equiv K - K_G$.
Namely, one has from Eqs.~(\ref{Rn},~\ref{lan1}):

\begin{equation}
\l_n \approx 1.20 \Delta K.
\label{lan2}
\end{equation}

\noindent
Eq.~(\ref{lan2}) is valid for convergents to the critical curve.
However, since this relation is based on Greene's formula
$R \propto K^q$~\cite{G79}, which is valid for all periodic orbits,
it is natural to assume that Eq.~(\ref{lan2}) is as well valid
in this general case. The periodic orbits densely fill phase space;
therefore, one can use Eq.~(\ref{lan2}) to predict local values of
the LLCE, $\l$, for all orbits residing in the chaotic layer not far
from its border. Hence

\begin{equation}
\l \propto \Delta K.
\label{la}
\end{equation}

\noindent
Thus the value of the LLCE is determined by that of the stochasticity
parameter $K$ of the approximating standard map.

The dependence of the transport time $T_r$ (equivalently
``recurrence'' or ``sticking'' time) near the border on $\Delta K$
is given by Chirikov's resonant theory of critical
phenomena~\cite[Section~4.3]{C90}. This dependence is derived as
follows. The recurrence time is of the same order as the time
$\tau_n$ of the transition from a scale $q_n$ with $n$ maximal for
a given recurrence, to the neigbouring one, since $\tau_n$ rapidly
diminishes with decreasing $n$~\cite{CS86}. The ``mean'' relation
$\Delta K \propto \rho$, where $\rho$ is the detuning $\vert r -
r_c \vert$ of the winding number $r$ with respect to that of the
critical curve, $r_c$, leads to $\Delta K \propto q_n^{-2}$. Such
a dependence is not sufficient to destroy principal critical
scales $q_n$; instead, narrow, $\sim q_n^{-4}$, chaotic layers are
formed between them. From the condition of the flux balance in
statistical equilibrium, one has $\tau_n \propto
q_n^4$~\cite{CS86,C90}. Since the recurrence time $T_r \sim
\tau_n$, and the dependence of $\Delta K$ on the winding number
detuning is set to be linear, one has

\begin{equation}
T_r \propto {\Delta K}^{-2}.
\label{tr}
\end{equation}

\noindent
Via Eqs.~(\ref{la},~\ref{tr}), one can then express the recurrence
time $T_r$ through the Lyapunov time $T_L$, which is the inverse of
the locally defined LLCE. One has finally

\begin{equation}
T_r \propto {T_L}^2.
\label{tt}
\end{equation}

\noindent Eq.~(\ref{tt}) is valid if, during the time of
measurement of the LLCE, the chaotic motion mostly takes place not
far from chaos border, i.e. in the ``sticking regime''. The
``sudden orbital change'' means escape from the border.

In the opposite limit of $K \gg 1$, i.e. for short recurrences,
the motion is completely diffusive, and the ``$T_L$~--- $T_r$''
relationship is not so simple. The relationship in the case of
diffusion was studied by Morbidelli and Froeschl\'e~\cite{MF96}.
They showed that it did not follow a uniform pattern; though power
laws with values of exponents in a wide range could be indeed
observed and explained.

The short-recurrence time (diffusive) part of the relationship seems
to be normally not observed in computations dealing with asteroidal
dynamics, due to a selection effect (the time of the measurement of
LCE cannot be too short). Besides, the diffusive stage can be simply
absent, e.g. as in the case of the SM with $\lambda \sim 1$.

The analysis above has been performed for the perturbed nonlinear
pendulum, which represents a system with one and a half degrees of
freedom (one degree of freedom plus dependence on time). Whether
the resulting formula has any relevance to systems with many
degrees of freedom? In such systems, the Arnol'd diffusion takes
place~\cite{C90}. This universal instability is not possible in
systems with the number of degrees of freedom less than or equal
to two. The situation in higher dimensions is therefore more
complicated. However, different resonances in multi-dimensional
systems generically have different strengths. According to
Chirikov's classification~\cite{C90}, the ``guiding'' resonance
may be chosen arbitrarily; this depends on the region of phase
space where the motion is considered. Remaining resonances are
considered as driving. The strongest one among the driving
resonances is the so-called ``layer'' resonance. It drives
transport across the chaotic layer of the guiding resonance. This
chaotic transport is faster than the Arnol'd diffusion driven by
remaining resonances, and can be described by the usual separatrix
map (see Ref.~\cite[p.~355]{C79}), retaining all statistical
properties of the latter. Therefore, one may expect that the
generic relationship ``$T_L$~--- $T_r$'' derived above in the
framework of the universal description of a perturbed one-degree
of freedom nonlinear resonance is generically valid for
multi-dimensional systems as well.

Another important problem concerns conditions for the emergence of
the generic relationship in numeric simulations. There are at
least two natural numeric selection effects, mentioned already in
Section~\ref{comp}, making the presence of the quadratic
relationship~(\ref{tt}) ubiquitous. These effects both concern the
procedure of measurement of the LLCE. First, the LLCE is measured
on a long enough time interval in order that its value were
``saturated'' (see Section~\ref{comp}). This eliminates small
recurrence times from consideration. Second, when relationships of
the kind ``$T_L$~--- $T_r$'' are constructed, the LLCE is not
measured normally on time intervals greater than $T_r$. It is
usually measured on some limited time interval, though long enough
for the LLCE to be saturated~(e.g.~\cite{LFM92}). This limitation
is usually justified by that the variation of the LLCE after its
saturation is slow.

Imposing a lower limit on the time of the measurement leads
to a situation that the impact of chaos border (or, equivalently,
the role of the sticking regime) may become prominent; imposing
the upper limit leads to that the LLCE corresponds to a particular
local domain of the chaotic region. These factors put in action
the generic relationship.

\section{Numeric examples}

I consider two examples of the emergence of the ``$T_L$~--- $T_r$''
relationship, one in a simpler and one in a computationally more
sophisticated problem. The statistical behaviour of a single
trajectory and that of a set of trajectories on a grid of starting
values are accordingly analyzed.

The first example straightforwardly concerns the SM
Eq.~(\ref{sm}). I build the dependence ``$\log T_L$~--- $\log
T_r$'', where $T_L$ is the inverse of the LLCE computed on a time
interval during which the trajectory stays at one side of the
chaotic layer; $T_r$ is the duration of this time interval. Time
is measured in iterations of the map. ``Staying at one side'' of
the layer means that the variable $w$ in Eq.~(\ref{sm}) has a
particular sign; when a trajectory crosses the central line of the
layer, the sign alternates; an orbit segment between crossings of
the central line forms a recurrence.

In computations, the SM~Eq.(\ref{sm}) was used in its equivalent
form~\cite{CS84}

\begin{eqnarray}
     y_{i+1} &=& y_i + \sin x_i, \nonumber \\
     x_{i+1} &=& x_i - \lambda \ln \vert y_{i+1} \vert + c
                                 \ \ \ (\mbox{mod } 2 \pi),
\label{sm1}
\end{eqnarray}

\noindent
where $y = w/W$, $x = \sigma + \pi$; the parameter
$c = \lambda \ln{32 / \vert W \vert}$ .

In Fig.~1, the dependence is shown for the values of the SM
parameters $\lambda = 3.22$, $c = 0$. These values are the same as
used in Ref.~\cite{CS84}. They correspond to a case of the
critical curve with the ``golden'' winding number, i.e. the
winding number equal to the golden mean $(\sqrt{5} - 1)/2$. The
latter is the irrational number furthest away from neighbouring
rationals~(see e.g. Ref.~\cite{LL92}). The effect of marginal
resonances is thus reduced to a minimum, and the quadratic nature
of the ``$T_L$~--- $T_r$'' relationship manifests itself most
clearly.

In captions to the figures, $n_{it}$ is the total number of
iterations in the computation run, $n_\mathrm{points}$ is the
number of points in the plot. Logarithmic scales in all figures
are decimal. Note that, in Fig.~1, the recurrences with duration
$T_r < 10$ are eliminated in order that the LLCE were saturated.
One can see that the dependence in Fig.~1 has a major pattern that
with large scatter but follows the straight line of the generic
relationship, as expected. I have built ``$T_L$~--- $T_r$''
dependences for various values of the SM parameters, not only for
the ``golden'' case. They always have the major pattern close to
the quadratic law. When marginal resonances are present, there are
disturbances, as expected.

Consider now an analytically more complicated problem, which
nevertheless has an established theoretical link to the SM paradigm.
The problem concerns the asteroidal motion in the 3/1
mean motion commensurability with Jupiter.
The following analysis is performed in the planar-elliptic restricted
three-body problem and is limited to asteroidal orbits with
eccentricities less than $0.4$. Trajectories
exhibiting the mode of jumps to very high eccentricities
$e \approx 1$ (cf. Ref.~\cite{FM93}) are not considered.

This asteroidal problem was shown~\cite{S96} to be
reducible, after averaging on the orbital time scale and
in certain regimes, to the SM with $\lambda \approx 1.4$.
Since the value of $\lambda$ is low, the diffusive stage is absent.
A manifestation of the behaviour well-known already in the case of
the SM was observed by Shevchenko and Scholl~\cite{SS96}
in the appearance of statistical distributions of duration of intervals
between eccentricity bursts of intermittent asteroidal orbits in the
3/1 Jovian resonance. The distributions in the tails followed the
power-law decay. This kind of dependence, according to
Chirikov~\cite{C90}, is immanent to trajectories' sticking
to chaos border.

Let us see how the relationship ``$\log T_L$~--- $\log T_r$''
looks like in the considered asteroidal problem.
The computations are performed with Wisdom's map~\cite{W83}.
The following notations are adopted henceforth:
$l$ and $l_J$ are mean longitudes of an asteroid and Jupiter;
$\varpi$ is the longitude of perihelion of the asteroid;
$a$ and $e$ are its semimajor axis and eccentricity.
The ratio of the mass of Jupiter to that of the Sun is set to be
equal to $1/1047.355$. Jupiter's perihelion is at the origin of
longitudes, i.e. $\varpi_J = 0$.

Consider orbits with starting values on the rectangular grid
$0.48025 \leq a_0 \leq 0.48200$, $0.005 \leq e_0 \leq 0.050$, with
the step in $a_0$ equal to $0.00005$ and that in $e_0$ equal to $0.005$.
For Jupiter, set the eccentricity $e_J = 0.048$;
the initial value of its mean longitude $\l_J$ set to be zero.
For an asteroid, set $l_0 = \pi$, $\varpi_0 = 0$. This choice of
$l_0$, $\varpi_0$ forms a representative plane of starting values of
the asteroidal motion~\cite{W83}: almost every orbit in the phase
space of the 3/1 Jovian resonance intersects this plane. The chosen
rectangle covers the domain of the chaotic motion at $e_0 \leq 0.05$
completely, as well as parts of the neighbouring space of regular
motion. With such a grid, one has $350$ orbits in all; $166$ among
them, those with $\log T_L < 5.3$, are chaotic.
The latter threshold value follows from an analysis of the bimodal
structure of the differential distribution of computed values of the
LLCE. One or two orbits with $\log T_L$ close to this value may be
controversial. In what follows, regular orbits are excluded.

Each trajectory, together with its LLCE, was computed on the time
interval $n_{it} = 10^7$ iterations of Wisdom's map~\cite{W83}
(one iteration equals to one Jupiter period), or less
if a burst of eccentricity was encountered. The burst was
considered to take place if the value $0.2$ of eccentricity
was surmounted. This provides a good empirical criterion in
the given range of starting eccentricities.

The resulting ``$\log T_L$~--- $\log T_r$'' relationship is shown
in Fig.~2. As in the case of the SM, one can see again that the
statistical dependence tentatively follows the generic relationship
expected for the motion near the separatrix of a nonlinear resonance.

An important feature of the observed dependence is that there
exists a group of chaotic orbits for which the recurrence time is
``infinite'', i.e. these orbits do not ever exhibit eccentricity
bursts, at least during the adopted time interval of computation.
They are displayed in Fig.~2 as points with $\log T_r$ at the
limiting value of $7$. Such orbits form a group located mostly at
$\log T_L = 4.0 \div 4.2$; thus they are definitely chaotic. A
closer analysis shows that these orbits have a very narrow
spectrum of winding numbers: the ratio $Q$ of frequencies of
rotation of angles used in the SM approximation of the relevant
motion~\cite{S96}, $\sigma \equiv 2 \varpi + l - 3 l_J$ and
$\kappa \equiv \varpi + l - 3 l_J$, lies within limits $4/3$ and
$3/2$ for these orbits; i.e. they are associated with the chaotic
layers around separatrices of the minor resonances $Q = 4/3$ and
$3/2$. The absence of eccentricity bursts simply means that these
resonances do not overlap with the integer one, $Q=1$, which is
responsible for the eccentricity bursts.

The existence of a chaotic asteroidal orbit without bursts was
already encountered by Milani and Nobili~\cite{MN92} in a study of
the asteroid Helga. Such phenomenon seems to be in an apparent
contradiction to the statistical law ``$T_L$~--- $T_r$'' as found
by~\cite{SFL90,LFM92,MLF94,LD93}. The example of the asteroidal
problem considered above shows that there is no contradiction; the
matter is in the definition of a ``sudden orbital change''. A
single definition should not be used when trajectories in a
statistical set belong to disconnected chaotic domains.

Concluding on the numerics, it is necessary to note the following.

\noindent
(1)~Exact values of the exponent $\beta$ of the power law fitting the
observed relationships in the examples considered are not presented here,
since the data cannot be straightforwardly corrected for selection effects.
The most important selection effect modifying the value of
the exponent calculated by the mean square fit is due to concentration
of the points to the lower part of the dependence, since long recurrences
are rare. Therefore the exponent value somewhat depends on the mode of
the data cutting at the lower time edge. Evidently, longer recurrences
should be taken with a greater weight,
presumably directly proportional to their duration.
Neglecting the weights would lead to a bias in computed values of the
exponent; author's numeric experiments with the SM show that the
value of $\beta$ somewhat diminishes as a rule. Maybe the small
deviation of the reported values of $\beta = 1.7~\div 1.9$
in Refs.~\cite{SFL90,LFM92,MLF94,LD93} from the theoretical
quadratic result is due to this selection effect.

\noindent
(2)~Whether the lower time limit of applicability of the theoretical
law~(\ref{tt}) is low enough for the examples considered?
Empirically, it seems to be the case. Indeed, the distribution of
duration of recurrences for the SM with $\lambda \sim 1$  transforms
into the algebraic law with the exponent $\approx -1.5$ (for the
integral distribution) characteristic of the sticking regime when
recurrences are just several iterations long. According to~\cite{C90},
the threshold value of $T_r$ for this transformation is
$\approx 0.3 \lambda^2$. Concerning the case of asteroidal trajectories
in the 3/1 Jovian resonance, the transformation of the distribution of
the inter-burst interval duration to the algebraic decay with the
exponent $\approx -1.5$ for integral distributions is often observed
already at $T_r \approx 10^5$ Jupiter periods~\cite{SS96}, i.e.
again the lower time limit for the theory's applicability is reasonably
small. These considerations are mostly empirical, of course; a universal
theoretical estimate for the lower time limit is still to be found.

\section{Conclusions}

In this paper, conditions for the emergence of a statistical
relationship between $T_r$, the recurrence time,
and $T_L$, the local Lyapunov time (the inverse of the locally defined
largest Lyapunov characteristic exponent, LLCE), were investigated
for the motion inside the chaotic layer around the separatrix of
a nonlinear resonance.

The generic relationship is shown to resemble the quadratic one.
The reasons leading to its emergence are very general. There are
two main factors. The first one is immanent to any Hamiltonian
system; this is the effect of trajectories' sticking to chaos
border. The second one is a natural selection effect arising in
the procedure of computation of LCE: though they are measured on
time intervals long enough for their values to be saturated, the
computation is normally stopped before or when sudden orbital
changes, which signal the end of the ``recurrence time'', take
place.

In order to check the validity of this theoretical relationship
straightforwardly, a statistical dependence between the duration of
a recurrence (the time which a chaotic trajectory stays at one side
of the chaotic layer) and the Lyapunov time (the inverse of the
numeric value of the LLCE measured on the time interval of the
recurrence) was constructed by means of computation with
the separatrix map~(\ref{sm}).
Besides, as a particular applied and numerically more complicated example,
a statistical dependence of the time of a sudden orbital change
(namely, the time of a burst of eccentricity) on the Lyapunov time was
constructed for chaotic asteroidal orbits in the 3/1 Jovian resonance
in the planar-elliptic three-body problem by means of
Wisdom's map~\cite{W83}. In both cases, the LLCE were measured
on a time interval equal to the recurrence time
(i.e. until a sudden orbital change), and the observed dependences
follow the generic relationship~(\ref{tt}), as expected.

The existence of the generic relationship~(\ref{tt}) tentatively
provides a theoretical explanation of the statistical dependences
of times of ``sudden orbital changes'' on Lyapunov times, found
by~\cite{SFL90,LFM92,MLF94,LD93} in numeric experiments in a
number of problems of celestial mechanics.

\smallskip
\begin{centerline}
{\bf Acknowledgement}
\end{centerline}
\smallskip

It is a pleasure to thank B.~V.~Chirikov for invaluable help, remarks
and discussions.

\begin{figure}[ht!]
\begin{center}
\includegraphics[width=12cm]{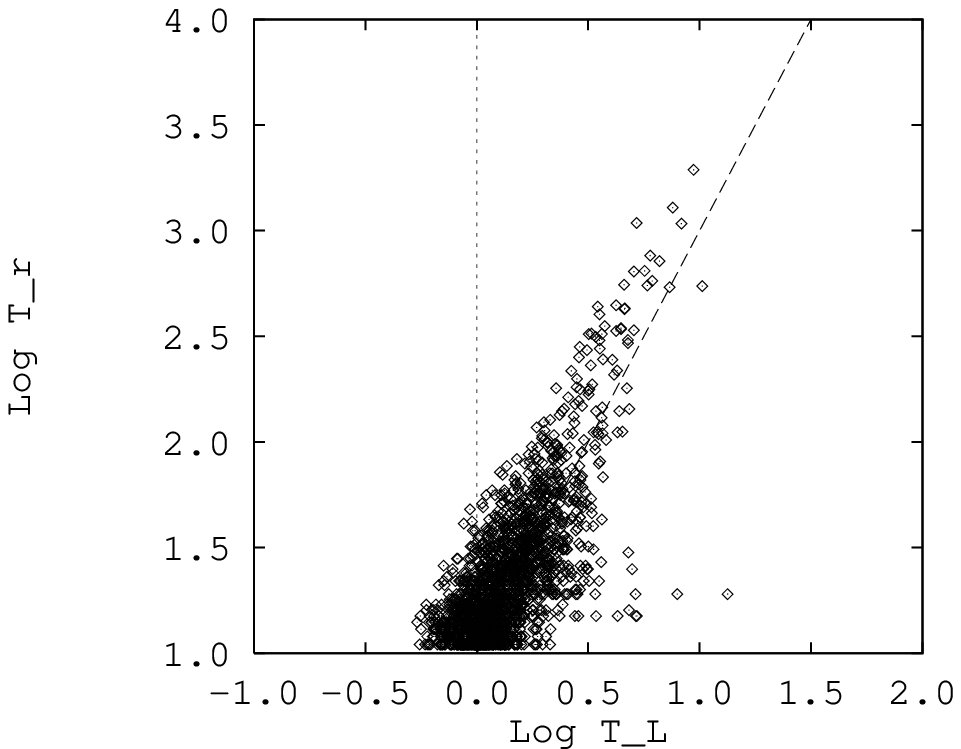}
\end{center}
\caption{\small The statistical dependence ``$\log T_L$~--- $\log
T_r$'' for the SM with parameters $\lambda = 3.22$, $c = 0$. The
winding number of the critical curve is ``golden''. $n_{it} =
10^5$, $n_\mathrm{points} = 1682$. Time is in iterations of the
map. The straight (dotted) line of the generic relationship is
shown for reference.} \label{fig1}
\end{figure}

\begin{figure}[ht!]
\begin{center}
\includegraphics[width=12cm]{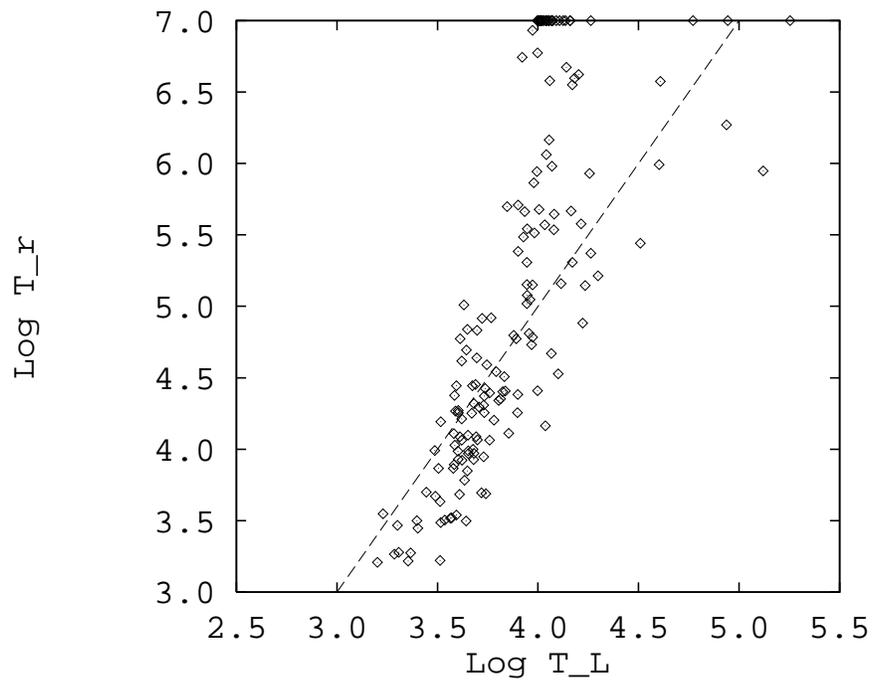}
\end{center}
\caption{\small The statistical dependence ``$\log T_L$~--- $\log
T_r$'' for chaotic asteroidal trajectories in the 3/1 Jovian
resonance, as described in the text. $n_\mathrm{points} = 166$.
Time is in Jupiter periods. The straight (dotted) line of the
generic relationship is shown for reference.} \label{fig2}
\end{figure}

\end{document}